# Over 600 V Lateral AlN-on-AlN Schottky Barrier Diodes with Ultra-Low Ideality Factor

Dinusha Herath Mudiyanselage, Dawei Wang, Ziyi He, Bingcheng Da, and Houqiang Fu, Member, IEEE

***Abstract*—This letter reports the demonstration of lateral AlN Schottky barrier diodes (SBDs) on single-crystal AlN substrates by metalorganic chemical vapor deposition (MOCVD) with an ultra-low ideality factor ($\eta$) of 1.65, a breakdown voltage (BV) of 640 V, and a record high normalized BV by the anode-to-cathode distance ($L_{AC}$). The homoepitaxially grown AlN epilayers had much lower defect densities and excellent surface morphology, and the AlN ohmic contacts also showed improvements. At forward bias, the devices exhibited ultra-low $\eta$ of 1.65 and high Schottky barrier height of 1.94 eV. The device current was dominated by thermionic emission, while most previously reported AlN SBDs suffered from defect-induced current with much higher $\eta$ of >4. Additionally, the devices also had excellent rectifying characteristics with ON/OFF ratios on the order of $10^7$ to $10^9$ and excellent thermal stability from 298 to 573 K. At reverse bias, the devices showed a high BV of 640 V and record-high normalized breakdown voltage (BV/$L_{AC}$) in lateral AlN SBDs. This work represents a big step towards high-performance ultra-wide bandgap AlN-based high-voltage and high-power devices.**

## I. Introduction

ALUMINUM Nitride is recently attracting research interests as an ultra-wide bandgap (UWBG) semiconductor for next-generation high-voltage and high-temperature electronics. This is mainly enabled by AlN's ultra-wide bandgap ($E_g$) of 6.2 eV and large critical breakdown field ($E_c$) of 12–15 MV/cm [1], [2], [3], [4], [5], [6], [7]. However, the development of AlN-based electronic devices is hindered by several challenges. First, the controllable *n*-type doping of AlN epilayers is still under development. Silicon (Si) is used as an *n*-type dopant in AlN with a high ionization energy of ~ 250 meV [8], [9]. At low doping levels, carbon impurities ($C_N$) and dislocations act as the compensators, while at high doping levels, $V_{Al}$ + nSi complexes reduce the free electron carrier concentration [10], [11], [12]. Due to this a "knee" behavior in free electron concentration was limited in Si-doped AlN epilayers with a maximum concentration of ~$2\times10^{15}$ cm$^{-3}$ [11]. Second, the formation of ohmic contacts is still challenging, and most metals have a Schottky-like behavior with AlN [4], [13], [14]. Furthermore, Schottky contacts to AlN with ideal Schottky barrier heights ($\varphi_b$) are difficult to achieve [15]. Recent studies on AlN Schottky barrier diodes (SBDs) have shown promising high-voltage electronic devices [16], [17], [18], [19], [20]. However, all these devices suffer from high ideality factors ($\eta$) >4.0, indicating the current transport mechanism deviates from the well-known thermionic emission (TE) theory. There may be several reasons for this. First, the current due to TE is very low due to high $\varphi_b$ (~2 eV for Ni Schottky contacts) [15], [21]. Second, the current through defects and dislocations may be present at low-bias conditions which reduces the TE current. At high bias, the current transport is limited by the series resistance of the AlN epilayers. Recently, Quiñones *et al.* [15] demonstrated low $\eta$ AlN SBDs ($\eta$ =1.5–1.7) with Ni Schottky contacts. They utilized large area rectangular ohmic contacts (~few mm$^2$) to suppress the dislocation/defect-induced current to enhance the TE current. However, the breakdown voltages (*BV*) of the devices were not studied, and the large device area may increase the total device cost, given the high cost of AlN substrates. In this work, we demonstrate AlN SBDs on single-crystal AlN substrates with an ultra-low $\eta$ of 1.65 and 640 V *BV*. This work represents a significant advancement in AlN power technology and opens the door to future development of high-performance UWBG AlN power electronics.

## II. Device Fabrication

The AlN epilayers were grown on single-crystal AlN substrates (dislocation density ~$10^3$ cm$^{-2}$) by metalorganic chemical vapor phase deposition (MOCVD). Trimethylaluminum (TMAl) and ammonia (NH$_3$) were used as the precursors, while silane (SiH$_4$) was the *n*-type dopant. The growth temperature and pressure are 1250 °C and 20 Torr, respectively. More growth details can be found elsewhere [18], [20]. The grown and fabricated device structure is shown in Fig. 1(a), which consisted of a 1-μm-thick AlN layer as a resistive buffer, a 200 nm highly Si-doped *n*-AlN layer, and a 2 nm UID GaN capping layer. The Si doping concentration in the *n*-AlN layer was $1\times10^{19}$ cm$^{-3}$. The GaN capping layer was used to prevent oxidation of the underlying AlN epilayers upon exposure to air, which could degrade device performance [22]. As shown in Fig. 1(c)-1(d), the homoepitaxially grown AlN

This work is supported as part of ULTRA, an Energy Frontier Research Center funded by the U.S. Department of Energy, Office of Science, Basic Energy Sciences under Award # DE-SC0021230. This work is also supported by the National Science Foundation (NSF) under Award No. ECCS-2338604.
Dinusha Herath Mudiyanselage, Dawei Wang, Ziyi He, Bingcheng Da, and Houqiang Fu are with the School of Electrical, Computer, and Energy Engineering, Arizona State University, Tempe, AZ 85287, USA (e-mail: houqiang@asu.edu).

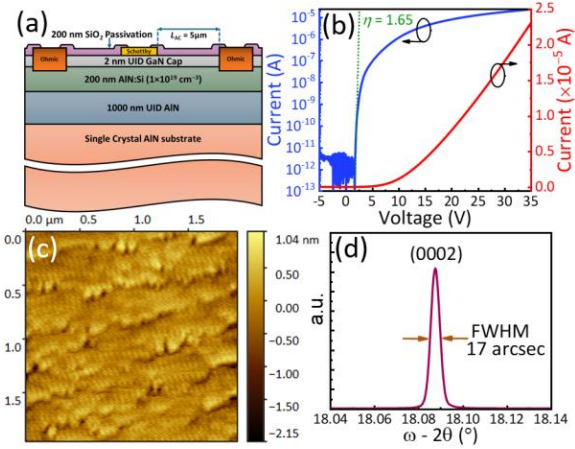

Fig. 1. (a) Schematic of the fabricated AlN SBD. (b) Forward current-voltage (I–V) characteristics on log and linear scales. (c) Atomic Force Microscopy (AFM) image and (d) (0002) Rocking curve of AlN epilayers grown on single-crystal AlN substrate.

epilayer has a smooth surface morphology with RMS roughness of ~0.4 nm by atomic force microscopy (AFM) and low dislocation density on the order of $10^4$ cm$^{-2}$ as measured by high-resolution X-ray diffraction (HRXRD). The use of single-crystal AlN substrates reduced the epilayer dislocation density by over three orders of magnitude compared with AlN epilayers on sapphire [18].

For the device fabrication, the sample first underwent a cleaning process involving acetone, isopropyl alcohol, and deionized water aided by ultrasonication, and hydrochloric acid to remove surface contaminations. The fabrication of AlN SBDs was performed using conventional optical photolithography and lift-off processes. Ohmic contacts were formed using Ti/Al/Ti/Au (25/100/25/50 nm) metal stacks deposited via electron beam (e-beam) deposition, followed by rapid thermal annealing (RTA) at 950 °C in N$_2$ for 30 seconds. The circular ohmic contact had a width of 100 μm. Simultaneously with ohmic contacts, 100×200 μm rectangular transfer length method (TLM) structures were fabricated to measure the AlN ohmic contact behavior. Ni/Au (25/125 nm) metal stacks were deposited via e-beam evaporation as the Schottky contacts. The Schottky contact had a diameter of 100 μm, and the cathode-to-anode distance, $L_{AC}$ was 5 μm. The devices were passivated using 200 nm SiO$_2$ by plasma-enhanced chemical vapor deposition (PECVD). Finally, the contact vias were opened using fluorine-based (SF$_6$) reactive ion etching (RIE). Electrical measurements were performed on a probe station equipped with a Keithly 4200 SCS semiconductor analyzer and a thermal chuck. Reverse I–V characteristics were measured using Keysight B1505A power device analyzer/curve tracer, and reverse breakdown measurements were conducted in insulating Fluorinert liquid FC-70 at room temperature.

### III. RESULTS AND DISCUSSIONS

Figure 1(b) shows the forward I–V characteristics of the AlN SBD on both log and linear scales. The $\eta$ and the $\varphi_b$ have been calculated from equations (1) and (2),

$$J = J_s \left[\exp\left(\frac{q(V-IR)}{\eta kT}\right) - 1\right] \quad (1)$$

$$J_s = A^* T^2 \exp\left(-\frac{q\varphi_b}{kT}\right) \quad (2)$$

where $k$, $T$, $R$, $A^*$, $J_s$, and $\varphi_b$ represent the Boltzmann constant, absolute temperature, series resistance, Richardson constant, reverse saturation current density, and Schottky barrier height, respectively. The $\varphi_b$ is usually replaced by effective Schottky barrier height $\varphi_{eff}$ when $\eta$ deviates from unity [19]. The device showed an ultra-low $\eta$ of 1.65 and a high $\varphi_{eff}$ of 1.94 eV. The $\eta$ of this work is among the record low values reported on AlN SBDs. This indicates that the current conduction is mainly due to TE and defect-induced current is minimized. Additionally, the $\varphi_{eff}$ of this device is also high (~1.9 eV), comparable to those of previously reported AlN high-voltage devices [16], [17], [18], [19], [20] that were predominantly governed by the defect-induced current transport.

Figure 2(a) shows AlN ohmic contact behavior at room temperature (RT). The current decreased as the gap between TLM pads ($d$) increased. The I–V behavior of the AlN ohmic contact is still non-linear due to the UWBG of AlN and low electron carrier concentration, which is commonly observed in AlN and high Al-content AlGaN [4], [19]. The I–V measurements were conducted from 298 (RT) to 573 K. Figure 2(b) shows the I–V behavior for $d$ = 10 μm for all the temperatures up to 573 K. The film sheet resistance (R$_s$) and contact resistance (R$_c$) were calculated as in Fig 2(c), and subsequently, the resistivity of AlN ($\rho$) and the contact resistivity ($\rho_c$) of ohmic contacts were extracted in Fig. 2(d). It should be noted that due to the nonlinear nature of the I–V measurements, the resistance is calculated at a certain voltage or current value [19], [23]. In this work, we calculate the resistance at 20 V, by dividing the corresponding current ($R = V/I$). We observed a reduced $\rho_c$ of $3.59 \times 10^{-2}$ Ωcm$^2$ at RT and a minimum value of $1.26 \times 10^{-3}$ Ωcm$^2$ at 473 K, which are comparable to high Al-content AlGaN [13], [14], [19]. Furthermore, the obtained $\rho_c$ was also lower than that of Si-ion implanted AlN on sapphire at RT and high temperatures, where the latter showed the lowest comparable $\rho_c$ of $4.0 \times 10^{-3}$ Ωcm$^2$ only at 1100 K [24].

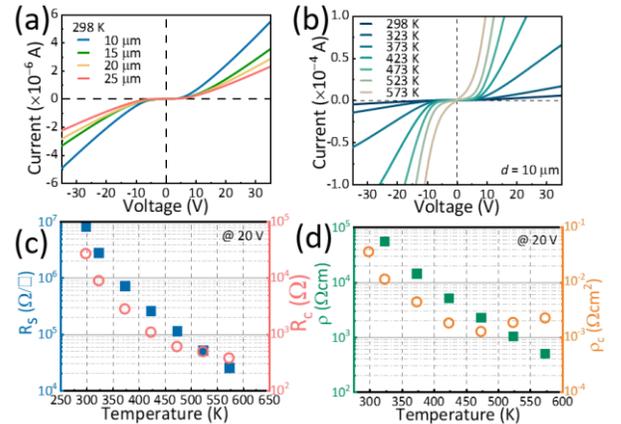

Fig. 2. (a) I–V characteristics of TLM structures at 298 K (RT) and (b) I–V characteristics of d = 10 μm TLM structure from 298–573 K. (c) Temperature-dependent sheet resistance (R$_s$) and contact resistance (R$_c$) and (d) AlN film resistivity ($\rho$) and contact resistivity ($\rho_c$) extracted from TLM measurements.

Figure 3(a) shows the temperature-dependent I–V characteristics of the AlN SBDs. The device showed a high ON/OFF ratio of $10^7$–$10^9$ as the temperature varies from RT to 573 K. Using the TE model, $\eta$ and $\varphi_{eff}$ were calculated at each temperature. The $\varphi_{eff}$ increased from 1.94 to 2.41 eV, and the $\eta$ decreased from 1.65 to 1.23 with increasing temperature. It should be noted that the $\varphi_{eff}$ is in good agreement with the predicted value for Ni Schottky contacts [15], [21]. It is worth noting that previous high-voltage AlN SBDs [16], [17], [18], [19], [20] failed to demonstrate high $\varphi_{eff}$ close to theoretical values due to deviation from the TE model. Figure 3(b) shows the temperature-dependence of $\varphi_{eff}$ and $\eta$ of the devices. This behavior can be attributed to an inhomogeneous metal/semiconductor interface with distributed high and low Schottky barrier regions. This behavior has also been commonly observed in previous reports [15], [18], [19], [20]. As temperature increases, electrons have sufficient energy to overcome higher Schottky barrier regions, leading to an increase in $\varphi_{eff}$.

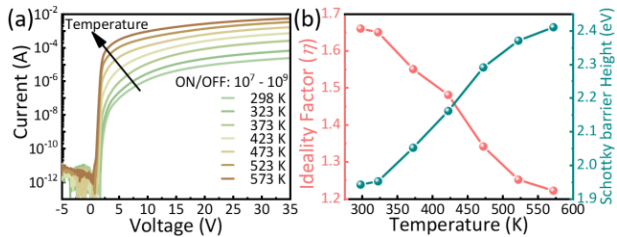

Fig. 3. (a) Temperature-dependent I–V characteristics of AlN SBDs. (b) Schottky barrier height ($\varphi_{eff}$) vs ideality factor ($\eta$).

Figure 4(a) shows the reverse breakdown measurements of the AlN SBDs under RT. The devices exhibited a *BV* of 640 V,

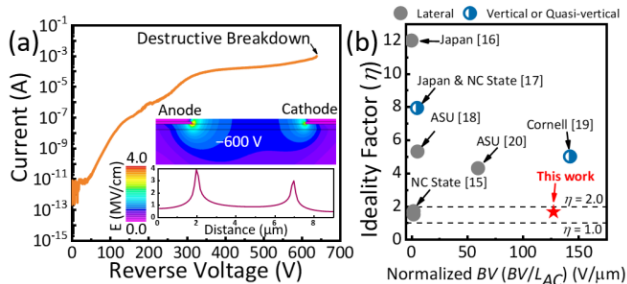

Fig. 4. (a) Breakdown characteristics of AlN SBD. The inset shows the TCAD simulation of the device under –600 V reverse bias. (b) Benchmark plot of ideality factor ($\eta$) and the normalized breakdown voltage (BV/$L_{AC}$) of the state-of-the-art AlN SBDs.

and the breakdown was destructive at the device edges due to the electric field crowding effect. TCAD simulations indicated the main cause of breakdown would be the crowded electric field under the anode edge with a peak field of 4 MV/cm. Effective edge termination and passivation with high k dielectrics are expected to further improve the device breakdown characteristics. Figure 4(b) shows the benchmark plot for state-of-the-art lateral AlN SBDs, where vertical and quasi-vertical devices are also included for reference. The normalized *BV* is calculated by dividing the device *BV* by the $L_{AC}$ (i.e., BV/$L_{AC}$) for comparison. Quiñones *et al*. [15] showed a low $\eta$ but a high *BV* was not reported. In other lateral high-voltage AlN SBDs, the $\eta$ was very high [16], [18], [20]. This work showed high *BV* and ultra-low $\eta$ simultaneously, which is one of the first demonstrations. The device in this work showed a record-high normalized *BV* in lateral AlN SBDs, which is also among the highest normalized *BV* ever reported in AlN SBDs. For vertical and quasi-vertical AlN SBDs, their *BV* is normalized by the drift layer thickness. Kinoshita *et al*. [17] showed a vertical AlN SBD with a *BV* of 770 and 150 µm epilayer by substrate removal. Maeda *et al*. [19] reported a quasi-vertical AlN SBD using AlGaN current spreading layer with 0.7 µm drift layer and *BV* of ~100 V. These vertical and quasi-vertical devices all showed large $\eta$ of >5.

## IV. Conclusion

In conclusion, we have successfully demonstrated MOCVD-grown AlN SBDs on single-crystal AlN substrates with an ultra-low $\eta$ of 1.65 and a *BV* of 640 V. The homoepitaxial growth of AlN epilayers showed excellent crystal quality and surface morphology, and the AlN ohmic contacts were also improved. The forward device characteristics are dominated by the TE model, resulting in ultra-low $\eta$ and high $\varphi_{eff}$. In addition, the device showed record-high normalized BV/$L_{AC}$ in lateral AlN SBDs. Furthermore, the device demonstrated stability at high temperatures up to 573 K. This work represents significant technological progress toward developing UWBG AlN-based high-voltage and high-power devices.